\documentclass[twocolumn,showpacs,floatfix]{revtex4}
\usepackage{latexsym}
\usepackage{graphics}
\usepackage{epsf}
\usepackage{epsfig}
\usepackage{graphicx}
\usepackage{dcolumn}
\usepackage{bm}
\usepackage{latexsym}
\usepackage{booktabs}
\usepackage{amssymb}
\usepackage[english]{babel}
\usepackage{hyperref}

\begin{document}

\noindent

\title{Entrainment of Coupled Oscillators on Regular Networks by Pacemakers}

\author{Filippo Radicchi\footnote{f.radicchi@iu-bremen.de} and Hildegard Meyer-Ortmanns\footnote{h.ortmanns@iu-bremen.de}}
\affiliation{School of Engineering and Science,
International University Bremen,\\
P.O.Box 750561, D-28725 Bremen, Germany}


\begin{abstract}
  We study Kuramoto oscillators, driven by one pacemaker, on
  $d$-dimensional regular topologies with  nearest neighbor interactions.
  We derive the analytical expressions for the common frequency in the case of
  phase-locked motion and for the critical frequency of the
  pacemaker, placed at an arbitrary position in the lattice,
  so that above the critical frequency no phase-locked motion is possible.
  We show that the mere change in topology from an open chain to a ring
  induces synchronization for a certain range of pacemaker
  frequencies and couplings, while keeping the other parameters fixed. Moreover
  we demonstrate numerically that the critical frequency of the pacemaker decreases
  as a power of the linear size of the lattice
  with an exponent equal to the dimension of the system.
  This leads in particular to the conclusion that for infinite-dimensional topologies
  the critical frequency for having entrainment
  decreases exponentially with the size of the system, or, more generally, with
  the depth of the network, that is the average distance of the oscillators from the pacemaker.
\end{abstract}


\maketitle

\section{Introduction}
Synchronization is an ubiquitous phenomenon, found in a variety of
natural systems like fireflies \cite{buck}, chirping crickets
\cite{walker}, or neural systems, but it is also utilized for
artificial systems of information science in order to enable a
well-coordinated behavior in time \cite{kanter}. As an important special case,
the coordinated behavior refers to similar or even identical units
like oscillators that are individually characterized by their
phases and amplitudes. A further reduction in the description was
proposed by Kuramoto for an ensemble of oscillators
\cite{kuraoriginal}, after Winfree had started with a first model
of coupled  oscillators \cite{winfree}. Within a perturbative
approach Kuramoto showed that for any system of weakly coupled and
nearly identical limit-cycle oscillators, the long-term dynamics
is described by differential equations just for the phases (not
for the amplitudes) with mutual interactions, depending on the
phase differences in a bounded form. It is this model, later named
after him, that nowadays plays the role of a paradigm for weakly
and continuously interacting oscillators.  In a large number of
succeeding publications the original Kuramoto model was
generalized in various directions, for a recent review see
\cite{kuramotoreview}. In particular the natural frequencies were
specialized in a way that one oscillator plays the role of a
pacemaker with frequency higher than the natural frequencies of
all other oscillators \cite{yamada}\cite{kori}. Pacemakers play an
important role for the formation of patterns in
Belousov-Zhabotinsky system \cite{zaikin}. Special families of
wave solutions of the phases arise as a consequence of dynamically
created pacemakers \cite{kuraoriginal} \cite{blasius}. Moreover
pacemakers are important for the functioning of the heart
\cite{winfree} and for the collective behavior of {\it
Dictyostelium discoideum} \cite{lee}, as well as for
large-scale ecosystems \cite{blasius1}. In \cite{weiss} it is shown that a single periodic pendulum oscillator can entrain or at least drastically influence the dynamics of all chaotic pendula on two-dimensional lattices. Furthermore in \cite{kori}
the role of pacemakers on complex topologies was analyzed in order
to understand the functioning of the neural network at the basis
of the circadian rhythm in mammals. \vspace{0.1cm}
\\
In this paper  we consider a system of  Kuramoto oscillators,
coupled with their nearest neighbors on various regular lattice
topologies, and driven by a pacemaker, placed at an arbitrary site
of the lattice (section \ref{sec:model}). In particular, we
analytically derive the common frequency  of phase-locked motion
in case of generic networks (in particular for $d$-dimensional
regular lattices with open or periodic boundaries) in section
\ref{sec:phase-locked}. Locked phases will be also called phase
entrainment throughout this paper. We also analytically derive the
upper bound on the absolute value of the ratio of the pacemaker's
frequency to the coupling strength in case of one-dimensional
regular lattices (section \ref{sec:onedim}). In section
\ref{sec:higherdim} we consider higher-dimensional lattices and
extend the results obtained for $d=1$ to any dimension $d \geq 2$
of the lattice. We find that the range of pacemaker frequencies
for which one obtains synchronization, the so-called entrainment
window, decreases with an inverse power of the linear size $N$ of
the lattice with an exponent given by the dimension $d$ of the
lattice. This leads to the conclusion that the entrainment window
of an infinite-dimensional network decreases exponentially with
its linear size $N$ if the pacemaker is asymmetrically coupled to
the other oscillators. This conclusion is supported by our
analysis of coupled oscillators on a Cayley-tree, a topology that
amounts to an infinite-dimensional regular lattice. These results
confirm the results recently obtained by Kori and Mikhailov
\cite{kori} for random network topologies. For random topologies
the entrainment window decays exponentially with the so-called
depth of the network, that is the average distance of all other
oscillators from the pacemaker. For our regular topologies, the
linear size of the networks for hypercubic lattices and the radius
of the Cayley tree are proportional to the network depth.
\section{The Model}
\label{sec:model}
The system is defined on a regular network. To each node $i$,
$i=0,\ldots, N$, we assign a limit-cycle oscillator, characterized
by its phase $\varphi_i$ that follows the dynamics
\begin{equation}
\dot{\varphi}_i = \omega + \delta_{i,s} \Delta \omega+  (1+ \delta_{i,s} \epsilon) \frac{K}{k_i} \sum_{j} A_{j,i} \sin{\left(\varphi_j-\varphi_i\right)}\;\;\; .
\label{def:model}
\end{equation}
$A$ is the adjacency matrix of the system ($A_{i,j}=A_{j,i}=1$ if
the nodes $i$ and $j$ are connected and $A_{i,j}=0$ otherwise), it
reflects the underlying topology of the network. Here only nearest
neighbors are coupled. Moreover, $k_i=\sum_{j}A_{j,i}$ is the
degree of the $i$-th node, it gives the total number of
connections of this node in the network. $\delta_{i,j}$ denotes
the Kronecker delta ( $\delta_{i,j}=1$ if $i=j$ and
$\delta_{i,j}=0$ if $i\neq j$ ). The oscillator at position $s$
represents the pacemaker. Its natural frequency differs by $\Delta
\omega$ with respect to the natural frequency $\omega$ of all
other oscillators. Without loss of generality we set $\omega =0$,
because system (\ref{def:model}) is invariant under the
transformation $\varphi_i \rightarrow \varphi_i+ \omega t\;\; , \;
\forall i$. Moreover the interaction of the pacemaker with the
other oscillators can be linearly tuned by the parameter $-1\leq
\epsilon \leq 0$. For $\epsilon = 0$ the pacemaker is on the same
footing as the other oscillators. For $\epsilon = -1$ its
interaction is asymmetric in the sense that the pacemaker
influences the other oscillators, but not {\it vice versa} (the
pacemaker acts like an external force). In natural systems both
extreme cases as well as intermediate couplings can be realized.
The constant $K>0$ parameterizes the coupling strength. The phases
of the $i$-th and $j$-th oscillators interact via the sine
function of their difference, as originally proposed by Kuramoto.
\section{Phase-locked Motion}
\label{sec:phase-locked}
We consider the conditions for having phase-locked motion , in
which the phase differences between any pair of oscillators remain
constant over time, after an initial short transient time. First
we calculate the frequency $\Omega$, in common to all oscillators
in the phase-locked state. Imposing the phase-locked condition
$\dot{\varphi}_i \equiv \Omega \;\;\; , \; \forall \; i=0,\ldots
,N$ to system (\ref{def:model}) and using the fact that the sine
is an odd function (see the Appendix A for details), we obtain

\begin{equation}
\Omega\;=\; \Delta \omega\;\frac{k_s}{\left(1+\epsilon\right)\sum_{i \neq s}
k_i + k_s} \; \; \; .
\label{eq:common_frequency}
\end{equation}
As long as $\Omega$ depends on the degree $k_s$ we see that the
common frequency $\Omega$ increases with the degree of the
pacemaker. On the other hand, when the network has a homogeneous
degree of connections, $k_i \equiv k \;\;\; , \forall \;
i=0,\ldots, N$, Eq.(\ref{eq:common_frequency}) takes the form
\begin{equation}
\Omega = \frac{\Delta \omega}{\left(1+\epsilon \right) N +1}
\;\;\;.\label{eq:common_frequency_ring}
\end{equation}
It should be noticed that in this case the common frequency does
not depend on the common degree $k$ of the network.
\\
In terms of the original parameterization of the model
(\ref{def:model}), the common frequency after synchronization is
$\Omega + \omega$. For the derivation of Eq.s
(\ref{eq:common_frequency}) and (\ref{eq:common_frequency_ring}) we
made only use of the odd parity of the coupling function. The
former results are still valid for any other odd coupling function
$f\left(\varphi_j-\varphi_j\right)$ which is $2\pi$-periodic and
bounded.
\section{One-dimensional lattice}
\label{sec:onedim}
\begin{figure}[ht]
\includegraphics*[width=0.47\textwidth]{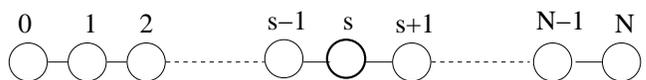}
\caption{One-dimensional lattice with $N+1$ sites labelled by their coordinates.}
\label{fig:chain}
\end{figure}
\subsection{Linear chain} Let us consider first the case of $N+1$
Kuramoto oscillators coupled along a chain (Figure
\ref{fig:chain}) with free boundary conditions  at the positions
$0$ and $N$ ($k_i=1$ if $i=0$ or $i=N$, $k_i=2$ otherwise), the
pacemaker located at position $s$, $0\leq s\leq N$. In this case
it is convenient to introduce the phase lag between nearest
neighbors $\theta_i=\varphi_i-\varphi_{i-1}$. Consider first the
pacemaker placed at position $0<s<N$. We start with the nearest
oscillators to the right of the pacemaker, placed at position
$i=s+1$
\[
\begin{array}{c}
\Omega=K/2\;\left[\;\sin{\left(-\theta_{s+1}\right)} + \sin{\left(\theta_{s+2}\right)}\;\right]
\\
\Rightarrow
\sin{\left(\theta_{s+2}\right)} = 2 \Omega / K + \sin{\left(\theta_{s+1}\right)}
 \end{array}\;.
\]
Moving again to the right, the equation of the $(s+2)$-th oscillator reads
\[
\begin{array}{c}
\Omega=K/2\;\left[\;\sin{\left(-\theta_{s+2}\right)} + \sin{\left(\theta_{s+3}\right)}\;\right]
\\
\Rightarrow
\sin{\left(\theta_{s+3}\right)} = 2 \Omega / K + \sin{\left(\theta_{s+2}\right)}
\\
\Rightarrow
\sin{\left(\theta_{s+3}\right)} = 4 \Omega / K + \sin{\left(\theta_{s+1}\right)}
 \end{array}\;.
\]
Iteratively, we can write for each $1\leq j \leq N-s$
\begin{equation}
\sin{\left(\theta_{s+j}\right)} = 2 \left(j-1\right) \frac{\Omega}{K}+  \sin{\left(\theta_{s+1}\right)}\;\;\;.
\label{eq:iter_right}
\end{equation}
In particular when $j=N-s$ we have
\[
\sin{\left(\theta_{N}\right)} = 2\left(N-s-1\right) \frac{\Omega}{K}+  \sin{\left(\theta_{s+1}\right)}\;\;\; ,
\]
but also at the boundary
\[
\Omega= K  \sin{\left(-\theta_N\right)}\;\;\;.
\]
From the last two equations we can simply determine the value of
$\sin{\left(\theta_{s+1}\right)}$ as function of $s$, $N$ and
$\Omega$. Substituting this value into Eq.(\ref{eq:iter_right}),
we obtain
\begin{equation}
\sin{\left(\theta_{s+j}\right)}\; = \;2(j-1) \frac{\Omega}{K}\;+\;
\left(2s-2N+1\right)\frac{\Omega}{K} \;. \label{eq:iter_right2}
\end{equation}
When $\Omega>0$ ($\Omega<0$), Eq.(\ref{eq:iter_right2}) is always
negative (positive) and has its minimum (maximum) value for $j=1$.
This means that when the pacemaker succeeds in "convincing" its
nearest neighbors to the right to adapt his frequency, all the
others to the right do the same. Now, the absolute value of the
critical threshold can be calculated by only using the fact that
the sine function is bounded ( $\left| \;
\sin{\left(\theta\right)} \; \right|\leq 1$ ) and using the
expression for $\Omega$ as a function of $\epsilon$ and $N$, as it
is derived in the Appendix A in Eq.(\ref{eq:common_frequency_app})
\begin{equation}
^R \left| \frac{\Delta\omega}{K}\right|_C =
\frac{\left(1+\epsilon\right)N-\epsilon}{2N-2s-1}\;\;\; .
\label{eq:critic_right}
\end{equation}
Eq.(\ref{eq:critic_right}) yields the bound for oscillators to the
right ($R$) of  $s$ to approach a phase-locked state. Following
the same procedure, but moving to the left of the pacemaker, we
find
\begin{equation}
^L \left|\frac{\Delta\omega}{K}\right|_C =
\frac{\left(1+\epsilon\right)N-\epsilon}{2s-1} \label{eq:critic_left}
\end{equation}
as bound for oscillators to the left ($L$) of the pacemaker to
synchronize in a phase-locked motion. Since we are interested in a
state with all oscillators of the chain being phase-entrained, we
need the stronger condition given by
\begin{equation}
\left|\frac{\Delta\omega}{K}\right|_C = \min{\left[ \; ^R
\left|\frac{\Delta\omega}{K}\right|_C \; , \;  ^L\left|
\frac{\Delta\omega}{K}\right|_C \; \right]}  \; \; . \label{eq:critic}
\end{equation}
For the pacemaker placed at the boundaries $s=0$ or $s=N$, using
Eqs.(\ref{eq:common_frequency}) and
(\ref{eq:common_frequency_app}), we obtain
\begin{equation}
\left| \frac{\Delta\omega}{K} \right|_C = \frac{\left(1+\epsilon\right)\left(2N-1\right)+1}{2N-1} \; \; . \label{eq:critic_bound}
\end{equation}
\begin{figure}[ht]
\includegraphics*[width=0.47\textwidth]{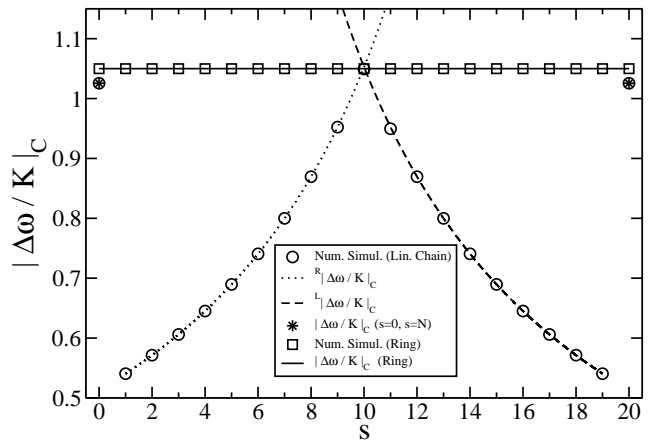}
\caption{Critical threshold $\left|\Delta \omega/K\right|_C$ as
function of the position $s$ of the pacemaker on a one-dimensional
lattice. For further details see the text.}
\label{fig:critic_one-dim}
\end{figure}
\subsection{Ring topology}
If we close the chain of $N+1$ oscillators to a ring, $k_i \equiv
2\; , \; \forall i$, the derivation of the upper bound on the
pacemaker's frequency $\left|\Delta\omega/K\right|_C$ proceeds in
analogy to that of Eq.(\ref{eq:critic_bound}).
Using Eq.(\ref{eq:common_frequency_ring}), the final result is
then given as
\begin{equation}
\left| \frac{\Delta\omega}{K}\right|_C\;=\;\frac{1}{N} + \left(1+\epsilon \right)\;\;\;.
\label{eq:critic_ring}
\end{equation}
In all former cases, for $N\to\infty$, the critical threshold
$\left|\Delta\omega / K\right|_C$ goes to $\left(1+\epsilon
\right)$ [or values proportional to  $\left(1+\epsilon \right)$].
Differently, the common frequency $\Omega$ goes to $\omega$ for $0
\geq \epsilon>-1$ and $N\to\infty$, while $\Omega$ goes to $\omega
+\Delta \omega$ for $\epsilon=-1$ and $N\to\infty$. Therefore the
symmetric coupling of the pacemaker to the rest of the system
favors the synchronizability of the system, while it can no longer
synchronize for a completely asymmetric coupling of the pacemaker
($\epsilon = -1$). This is plausible as it must be easier for $N$
oscillators to convince one pacemaker to follow them (case
$\epsilon>-1$) than the opposite case, in which the pacemaker must
convince $N$ oscillators to follow it (case $\epsilon=-1$). For
increasing system size, the latter case becomes impossible, while
the former is still possible.
\\
\subsection{Topological switch to synchronization}
\begin{figure}[ht]
\includegraphics*[width=0.47\textwidth]{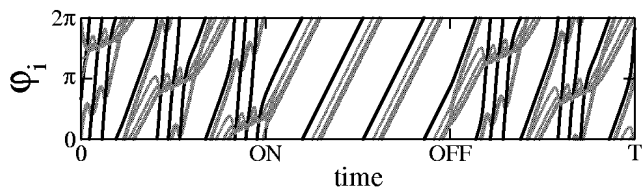}
\caption{Phase portrait of a system of Kuramoto oscillators on a
one-dimensional lattice. This figure shows how it is possible to
switch ``ON'' and ``OFF'' synchronization via a simple topological
change of the system, passing from a linear chain to a ring and
vice versa. A more detailed description is given in the
text.}\label{fig:switch}
\end{figure}

The numerical results of this paper are obtained by integrating
the set of Eq.s(\ref{def:model}) with the Runge-Kutta method of
fourth order ($dt=0.1$). The numerical value of the critical
threshold $\left| \Delta \omega / K \right|_C$ is evaluated with
an accuracy of $5 \cdot 10^{-5}$. As one knows from
\cite{kuraoriginal}, in case of regular topologies, stable
solutions with different winding numbers are possible. To avoid different winding numbers, we always
choose homogeneous initial conditions (also the distribution of {\it phases} would be different in particular in the synchronized state).
\\
We summarize the results obtained so far in Figure
\ref{fig:critic_one-dim}. The analytical results for
$\left|\Delta\omega/K\right|_C$ are represented by lines for the
open chain [dotted line from Eq.(\ref{eq:critic_right}) and dashed
line from Eq.(\ref{eq:critic_left})], for the ring [full line from
Eq.(\ref{eq:critic_ring})] and by crosses [Eq.(\ref{eq:critic_bound})] in case of an open chain with the
pacemaker at the boundaries, while the circles (open chain) and
squares (ring) represent numerical data that reproduce the
analytical predictions within the numerical accuracy. All results
are obtained for $N=20$  and $\epsilon=0$, they are plotted as a
function of the pacemaker's position $s$ that only matters in case
of the open chain. The horizontal line obviously refers to the
ring, the two branches (left and right), obtained for the chain,
cross this line when the pacemaker is placed at $s=10$ in the middle of the chain. When the pacemaker is located at the
boundaries $s=0$ and $s=20$, we obtain two isolated data points
close to the horizontal line.
\\
Let us imagine that for given $N$ and $\epsilon$ the absolute
value of the pacemaker's frequency $\left|\Delta\omega\right|$ and
the coupling $K$ are specified out of a range, such that the ratio
is too large to allow for phase-locked motion on a chain, but
small enough to allow the phase-entrainment on a ring. It is then
the mere closure of the open chain to a ring that leads from
non-synchronized to synchronized oscillators  with phase-locked
motion. Therefore, for a whole range of ratios
$\left|\Delta\omega/K\right|$, no finetuning is needed to switch
to a synchronized state, but just a simple change in topology, the
closure of a chain  to a ring. Because this closure may be much easier feasible
in real systems than a fine tuning of parameters to achieve synchronization, we believe
that this mechanism is realized in natural systems and should be utilized in artificial ones.
In our numerical integration we
simulated  such a switch and plot the phase portrait in Figure
\ref{fig:switch}. The phases $\varphi_i$ as function of time are
always projected to the interval $\left[0,2\pi \right)$: we use a
thick black line for the phase of the pacemaker and thin dark-grey
lines for the other oscillators. In this concrete numerical
simulation with $T \cdot  dt =4000$ integration steps altogether,
we analyzed a one-dimensional lattice of Kuramoto oscillators with
$N=6$, $s=2$, $\Delta\omega/K=1$ and $\epsilon=0$. In the time
interval from $0$ to ``ON'' ($T/3$) we see a phase evolution with
different slopes and varying with time. Moreover, the pacemaker
and the left part of the system (oscillators $i=0,1$) have larger
frequencies than the right part of the system (oscillators
$i=3,4,5,6$). At the instant ``ON'' we close the chain, passing to
a ring topology, the system almost instantaneously reaches a
phase-locked motion (all phases moving with the same and constant
frequency). At time ``OFF'' ($2T/3$) we open the ring, again, the
system then behaves similarly to the first phase, i.e. for $t\in
[0,ON)$.
Furthermore it should be noticed from Figure
\ref{fig:critic_one-dim} that it is also favorable to put the
pacemaker at the boundaries of an open chain to facilitate
synchronization. For $d=1$ the pacemaker then has to entrain only
one rather than two nearest neighbors so that the range of allowed
frequencies $\left|\Delta\omega\right|$ increases.
\section{Higher dimensions}
\label{sec:higherdim}
\begin{figure}[ht]
\includegraphics*[width=0.47\textwidth]{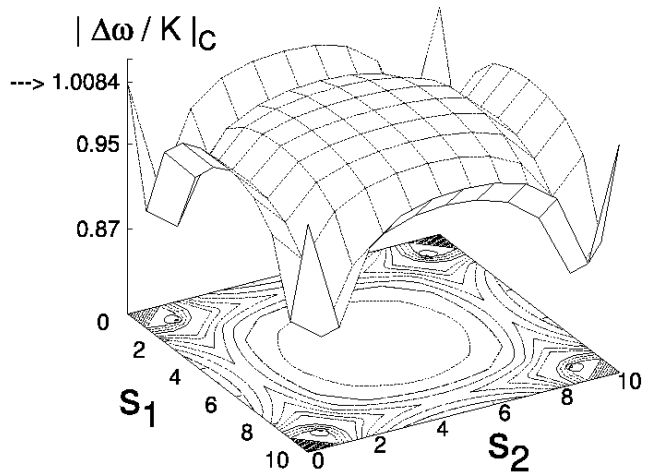}
\caption{Same as Figure \ref{fig:critic_one-dim}, but in $d=2$
dimensions. The linear size of the system is $N=10$. The arrow
indicates the value of the critical threshold for periodic
boundary conditions.}\label{fig:2dim}
\end{figure}

\begin{figure}[ht]
\includegraphics*[width=0.47\textwidth]{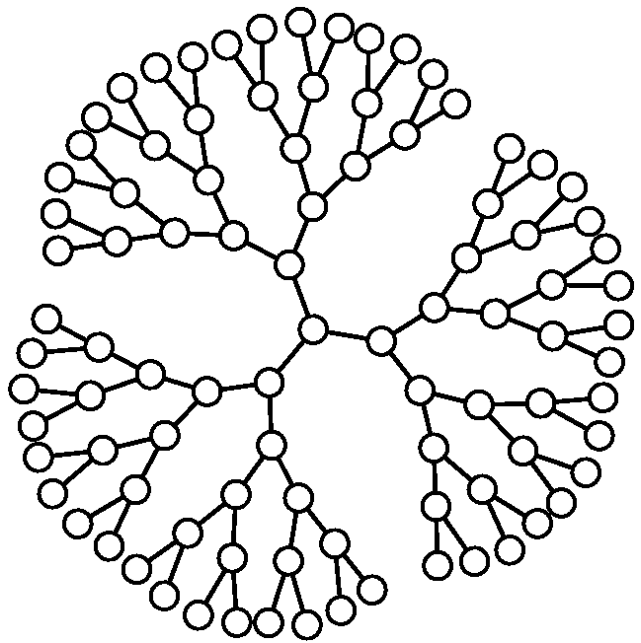}
\caption{Cayley tree with $z=3$ branches and radius $R=5$.}\label{fig:cayley}
\end{figure}

\begin{figure}[ht]
\includegraphics*[width=0.47\textwidth]{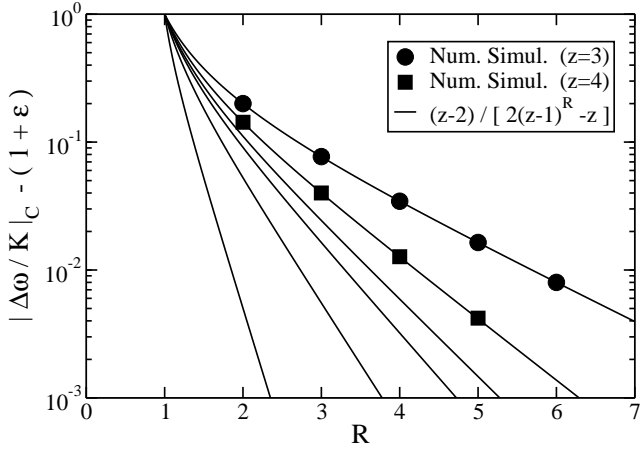}
\caption{Entrainment window for Cayley trees as a function of the
radius $R$. Numerical results (full dots and squares) refer to $z=
3$ and $4$. The full lines correspond to plots of
$\left|\Delta\omega / K \right|_C - \left(1+\epsilon\right)$ , as defined in Eq.(\ref{eq:critic_cayley}), for
different coordination numbers $z$, from up to down $z= 3, 4, 5,
6, 10$ and $ 100$, respectively.}\label{fig:higherdim}
\end{figure}

All of our results obtained so far extend qualitatively to higher
dimensions $d$, when system (\ref{def:model}) is placed on a
hypercubic lattice with $(N_j+1)$ oscillators in each direction
$j$, so that we have an ensemble of $\Pi_{j=1}^d(N_j+1)$ Kuramoto
oscillators, where the $i$-th oscillator's position is labelled by
a $d$-dimensional vector $\vec{i}$, with $0\leq i_j\leq N_j\; , \;
\forall \;j=1,\ldots ,d$. If the condition for having a
phase-locked motion is satisfied, the system of oscillators
reaches a common frequency still given by
Eq.(\ref{eq:common_frequency}). Such condition now is satisfied
for $\left|\Delta\omega/K\right| \leq \left|
\Delta\omega/K\right|_C$ , the critical ratio for the pacemaker's
frequency at position $\vec{s}=(s_1,\ldots ,s_d)$[see Figure
\ref{fig:2dim}].
In the simplest case when the lattice has $(N+1)$ sites in each
direction, we can qualitatively extend all the previous results,
obtained so far for $d=1$, to any dimension $d\geq 2$. For
example, the ``closure'' of the open boundaries of the lattice to
a torus in $d$ dimensions favors synchronization of the system. We
checked this numerically for $d=2$ and $N=10$ [Figure
\ref{fig:2dim}]. Except for the central node at $\vec{s}=(5,5)$,
the critical threshold in case of open boundary conditions lies
always below that for periodic boundary conditions. Moreover, it
is natural to assume that the former results, obtained for
one-dimensional lattices, in the case of periodic boundary
conditions extend to $d$-dimensional lattices by replacing $N$ in
Eq.(\ref{eq:critic_ring}) by $(N+1)^d-1$
\begin{equation}
\left|\frac{\Delta\omega}{K}\right|_C =
\frac{1}{\left(N+1\right)^d-1} + \left(1 + \epsilon \right)\;\;\;.
\label{eq:critic_ring_higher}
\end{equation}
\begin{figure}[ht]
\includegraphics*[width=0.47\textwidth]{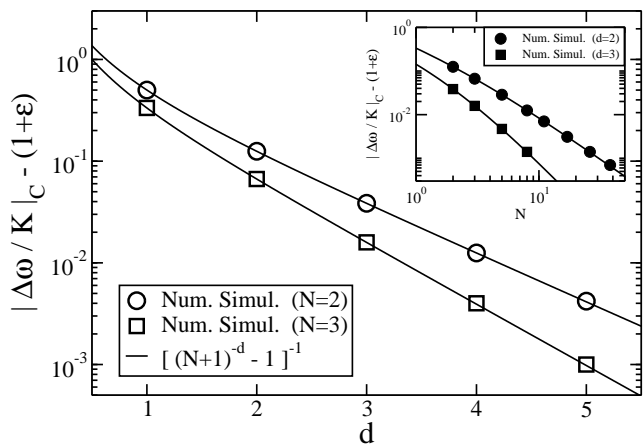}
\caption{Entrainment window for higher dimensional lattices. We
plot the critical threshold $\left| \Delta\omega /K
\right|_C-(1+\epsilon)$ that is independent on $\epsilon$ as it is seen from Eq.(\ref{eq:critic_ring_higher}).
The main plot shows the dependence of the entrainment window on
the dimension $d$, while the inset shows its dependence on the
linear size $N$ of the lattice. Numerical results are plotted as
symbols. They fit perfectly with our predictions according to
Eq.(\ref{eq:critic_ring_higher}).}\label{fig:dim}
\end{figure}
For $\epsilon=-1$ and fixed $K$, the entrainment window
$\Delta\omega_c$ decreases exponentially with the dimension $d$ of
the lattice and as a power of the linear size of the lattice with
exponent $d$. This conjecture is supported by the numerical
results as documented in Figure \ref{fig:dim}. For this reason we
expect that for infinite-dimensional systems ($d \to \infty$) the
entrainment window decreases exponentially with the linear size of
the system if the pacemaker is asymmetrically coupled with
$\epsilon = -1$.
\\
In order to verify this conjecture, we study a system of
limit-cycle oscillators placed on a Cayley tree [Figure
\ref{fig:cayley}].
Cayley trees have $z$ branches to nearest neighbors at each node,
apart from nodes in the outermost shell, where the number of
nearest neighbors is $z-1$. Cayley trees are infinite-dimensional
objects in the sense that the surface of the system (i.e. the
number of nodes in the outermost shell at distance $R$ from the
the center) is proportional to the volume of the system (i.e. the
total number of nodes in the tree).  The set of evolution
equations is  still given by Eq.(\ref{def:model}), with the
adjacency matrix $A$ specialized to the underlying Cayley tree
topology. For simplicity, we consider only the case of the
pacemaker placed at the center of the tree ($s=0$).
\\
When the number of branches per nodes $z$ is two, the Cayley tree
becomes a linear chain, with $N=2R$ and the pacemaker placed at
position $s=N/2=R$. Obviously for $z=2$ we find the same results
as given in the previous section (see the Appendix B).
\\
The infinite dimensionality of the tree shows up for $z>2$. We
obtain
\begin{equation}
\Omega  =  \Delta \omega \; \frac{z-2}{z-2   +   (1+\epsilon) \left[ 2(z-1)^R -  z\right]}
\label{eq:common_frequency_cayley}
\end{equation}
as common frequency for the phase-locked motion and
\begin{equation}
\left|\frac{\Delta\omega}{K}\right|_C \; = \;
\frac{z-2}{2\left(z-1 \right)^R -z} + \left(1+\epsilon \right)
\label{eq:critic_cayley}
\end{equation}
as critical ratio for obtaining the phase-locked motion. Both Eq.s
(\ref{eq:common_frequency_cayley}) and (\ref{eq:critic_cayley})
are valid for $R \geq 2$. In particular
Eq.(\ref{eq:critic_cayley}) tell us that for $\epsilon = -1$ the
entrainment window \emph{decreases exponentially} with the radius
$R$ of the Cayley tree (see Figure \ref{fig:higherdim}). For large
$z \geq 2$ it is easily seen that the radius $R$ gets
proportional to the depth $D$ of the network (in the limit $z \gg 2$ one finds $D \simeq R$).
These results nicely confirm those recently obtained by Kori and
Mikhailov \cite{kori}, who found that in infinite-dimensional
systems such as random networks and small-world networks with a
high number of rewired edges, the entrainment window decreases
exponentially with the depth of the network.
In order to connect their results to ours, we note that not only
the radius $R$ of the Cayley tree, but also the linear size $N$ of
the hypercubic lattices are proportional to the depth of these
regular networks.
\section{Summary and conclusions}
\label{sec:conclusion}
Entrainment of Kuramoto oscillators, coupled on regular lattices
via a pacemaker is possible for large system sizes ($N \to
\infty$) only in the case of symmetric ($\epsilon =0$) couplings
of the pacemaker. As the pacemaker coupling becomes asymmetric
($-1 \leq \epsilon < 0$), synchronization becomes more and more
difficult, and impossible for large system sizes and
$\epsilon=-1$. However, we are not only interested in the
``thermodynamic'' limit. For finite $N$ and $-1\leq\epsilon\leq
0$, we find that for a whole range of ratios
$\left|\Delta\omega/K\right|$ it is possible to induce
synchronization by a mere closure of a chain to a ring. The
sensitive dependence of synchronization on the topology in a
certain range of parameters may be exploited in artificial
networks and is -very likely- already utilized in natural systems,
in which a switch to a synchronized state should be easily
feasible (although we are currently not aware of a concrete
example from biological systems). If the pacemaker is coupled
symmetrically to the other oscillators ($\epsilon = 0$), the
entrainment window stays finite in the large-$N$ limit, but the
common frequency approaches zero for $N\to\infty$. In the other
extreme case, if the pacemaker is coupled asymmetrically to the
rest of the system ($\epsilon=-1$), our main result is that the
entrainment window decreases as a power of the depth of the
network with the dimension $d$ in the exponent. Extrapolating this
behavior to arbitrary dimension $d$, we see that one of the
reasons of the exponentially fast "closure" of the entrainment
window in complex network topologies \cite{kori} is their
effective infinite dimensionality.
\begin{appendix}
\section{Common frequency}
\label{app:common_frequency} Imposing the phase-locked condition,
setting $\omega=0$ and dividing by $K/k_i$,$\forall \;
i=0,\ldots,N$, system (\ref{def:model}) takes the form
\begin{equation}
\frac{k_i}{K}\Omega \; = \;  \delta_{i,s}\frac{k_i}{K} \Delta\omega \; +\; \left(1+\delta_{i,s}\epsilon\right) \Omega_i\;\;\; ,
\label{def:model_app}
\end{equation}
where
\[
\Omega_i\; := \; \sum_{j\neq i} \sin{\left(\varphi_j-\varphi_i\right)} \;\;\;.
\]
{}From the odd parity of the sine function it follows that
\[
\sum_i \Omega_i=0 \;\;\; \textrm{and} \;\;\; \sum_{i\neq g} \Omega_i = -\Omega_g\;\;\;.
\]
Summing (\ref{def:model_app}) over all $i$ except $i=s$ we obtain
\[
\begin{array}{l}
\sum_{i\neq s} \Omega \frac{k_i}{K} = \sum_{i\neq s}
\delta_{i,s}\frac{k_i\Delta \omega}{K} +
\; \sum_{i \neq s} \left(1+\delta_{i,s}\epsilon\right) \Omega_i
\\
\\
\Rightarrow \frac{\Omega}{K} \sum_{i \neq s} k_i = - \Omega_s
\end{array}
\; ,
\]
while summing (\ref{def:model_app}) over all $i$ except $i=j$, where $j\neq s$, we obtain
\[
\begin{array}{l}
\sum_{i\neq j} \Omega \frac{k_i}{K}= \sum_{i\neq j}
\delta_{i,s} \frac{k_i\Delta\omega}{K}+\sum_{i\neq j}
(1+\delta_{i,s}\epsilon) \Omega_i
\\
\\
\Rightarrow \ \frac{\Omega}{K}\sum_{i\neq j} k_i =
\frac{k_s\Delta\omega}{K}- \Omega_j + \epsilon \Omega_s
\\
\\
\Rightarrow \ \frac{\Omega}{K}\sum_{j\neq s}\sum_{i\neq j} k_i
= \sum_{j\neq s}\frac{k_s\Delta\omega}{K}- \sum_{j\neq s} \Omega_j
+ \sum_{j \neq s} \epsilon \Omega_s
\\
\\
\Rightarrow \ \frac{\Omega}{K}\sum_{j\neq s}\sum_{i\neq j} k_i
= \frac{Nk_s\Delta\omega}{K}+(1+\epsilon N)\Omega_s
\end{array}
\;.
\]
It should be noticed that
\[
\sum_{j\neq s}\sum_{i\neq j}k_i = Nk_s+\left(N-1\right)\sum_{i\neq s}k_i\;\;\;,
\]
so that we can write
\[
\Omega\left[\left(1+\epsilon\right)\sum_{i\neq s} k_i +
k_s   \right] =
k_s\Delta\omega\; \; \; ,
\]
{} from which Eq.(\ref{eq:common_frequency}) is implied.
\\
In the case of a one-dimensional lattice with open boundary
conditions, we have $k_0=1=k_N$ and $k_j=2 \;\; , \; \forall j\neq
0,N$, so that
\[
\sum_{i \neq s}k_i\; =
\left\{
\begin{array}{ll}
2N-1 & \textrm{ , if } s=0 \textrm{ or } s=N
\\
2\left(N-1\right) & \textrm{ , otherwise}
\end{array}
\right.
\;\;\;.
\]
 The common frequency of Eq.(\ref{eq:common_frequency}) can be written as
\begin{equation}
\Omega \; =\; \Delta \omega
\left\{
\begin{array}{ll}
\frac{1}{\left(1+\epsilon\right)\left(2N-1\right)+1} & \textrm{ , if } s=0 \textrm{ or } s=N\\
\frac{1}{\left(1+\epsilon\right)\left(N-1\right)+1} & \textrm{ ,
otherwise}
\end{array}
\right. \;. \label{eq:common_frequency_app}
\end{equation}
\section{Common frequency and critical threshold for Cayley trees}
\label{app:cayley} Consider a node in the outermost shell, i.e. at
distance $R$ from the pacemaker, reached along the path
$\vec{s}_R=(s_1,s_2,s_3,\ldots,s_{R-1},s_R)$ from the central
node. Here $s_1=1,\ldots,z$ , while $s_i=1,\ldots,z-1$ , $\forall
i =2,\dots,R$ [see Figure \ref{fig:cayley}] label the choice of
branch along the path. For this node we can write Eq.s
(\ref{def:model}) as
\[
\begin{array}{l}
\dot{\varphi}_{\vec{s}_R} =\Omega = \\
=K \sin{\left(\varphi_{\vec{s}_{R-1}} -
\varphi_{\vec{s}_R}\right)}
\\
\Rightarrow \ \sin{\left(\varphi_{\vec{s}_{R-1}} -
\varphi_{\vec{s}_R}\right)} = \frac{\Omega}{K}
\end{array}
\; ,
\]
from which we can see that all  oscillators at distance $R$,
independently on the path along which they are reached from the
center, satisfy the same equation.
\\
Proceeding in an analogous way as before, we find for the shell
$R-1$
\[
\begin{array}{l}
\dot{\varphi}_{\vec{s}_{R-1}} = \Omega =
\\
\\
= \frac{K}{z} \Big[ \sum_{s_R=1}^{z-1}  \
\sin{\left(\varphi_{\vec{s}_R} -
\varphi_{\vec{s}_{R-1}}\right)} +
\\
\\
\quad + \sin{\left(\varphi_{\vec{s}_{R-2}} -
\varphi_{\vec{s}_{R-1}}\right)} \Big] =
\\
\\
= \frac{K}{z} \Big[ - (z-1) \frac{\Omega}{K} +
\sin{\left(\varphi_{\vec{s}_{R-2}} -
\varphi_{\vec{s}_{R-1}}\right)} \Big]
\\
\\
\Rightarrow \ \sin{\left(\varphi_{\vec{s}_{R-2}} -
\varphi_{\vec{s}_{R-1}}\right)} = \frac{\Omega}{K} [ z + (z-1)]\;.
\end{array}
\]
Furthermore, for the shell $R-2$ we obtain
\[
\begin{array}{l}
\dot{\varphi}_{\vec{s}_{R-2}} = \Omega =
\\
\\
\frac{K}{z} \Big[ \sum_{s_{R-1}=1}^{z-1}  \
\sin{\left(\varphi_{\vec{s}_{R-1}} -
\varphi_{\vec{s}_{R-2}}\right)} +
\\
\\
\quad + \sin{\left(\varphi_{\vec{s}_{R-3}} -
\varphi_{\vec{s}_{R-2}}\right)} \Big] =
\\
\\
= \frac{K}{z} \Big[ - (z-1) \frac{\Omega}{K} [z+(z-1)] +
\sin{\left(\varphi_{\vec{s}_{R-3}} -
\varphi_{\vec{s}_{R-2}}\right)} \Big]
\\
\\
\Rightarrow \ \sin{\left(\varphi_{\vec{s}_{R-3}} -
\varphi_{\vec{s}_{R-2}}\right)} = \frac{\Omega}{K}
\left[ z + z(z-1)+(z-1)^2\right]
\end{array}
\]
until we arrive for $R-(t+1)$ at
\[
\sin{\left(\varphi_{\vec{s}_{R-(t+1)}} -
\varphi_{\vec{s}_{R-t}}\right)} =
\frac{\Omega}{K}  \left[  z  \sum_{q=0}^{t-1} (z-1)^q +
(z-1)^t \right]\; \; \; \; .
\]
For $t=R-1$, we have
\[
\sin{(\varphi_{\vec{s}_0} - \varphi_{\vec{s}_1})} \; = \; \frac{\Omega}{K}   \left[  z
\sum_{q=0}^{R-2}  (z-1)^q   +   (z-1)^{R-1}  \right]  \; \; \; \; ,
\]
but also, at the center of the Cayley tree,
\[
\begin{array}{l}
\Omega = \dot{\varphi}_{\vec{s}_0} =
\\
\\
= \Delta \omega +  \left(1+ \epsilon \right) \frac{K}{z}  \sum_{s_1=1}^z
 \sin{\left(\varphi_{\vec{s}_1} - \varphi_{\vec{s}_0}\right)} =
\\
\\
=
 \Delta \omega -  \left(1+\epsilon\right) \Omega  \left[  z  \sum_{q=0}^{R-2}  (z-1)^q   +   (z-1)^{R-1}  \right]
\end{array}
\;,
\]
{} from which
\begin{equation}
\Omega \; = \; \Delta \omega \; \left[  1  + \left(1+\epsilon\right) \left( z\sum_{q=0}^{R-2}
(z-1)^q  +  (z-1)^{R-1}\right)  \right]^{-1} \; \; \; .
\label{eq:common_frequency_cayley_app}
\end{equation}
For $z=2$ the Cayley tree reduces to a linear chain with $N+1$
oscillators ($N=2R$), open boundary conditions and the pacemaker
placed at position $s=N/2$. The common frequency given in
Eq.(\ref{eq:common_frequency_cayley_app}) becomes the same as in
Eq.(\ref{eq:common_frequency_app}).
\\
For $z>2$ we can rewrite the truncated geometric series as
\begin{equation}
\sum_{q=0}^{R-2} \left(z-1\right)^q = \frac{\left(z-1\right)^{R-1}-1}{z-2}
\label{eq:truncated_series}
\end{equation}
and obtain after some algebra
Eq.(\ref{eq:common_frequency_cayley}).
\\
As we have seen,  all  oscillators at the same distance $r$ from
the pacemaker satisfy the same equation. We can write
\begin{equation}
\sin{\left(\varphi_{r} - \varphi_{r-1}\right)}  =   -
\frac{\Omega}{K} \left[ z  \sum_{q=0}^{R-r-1}  (z-1)^q  +  (z-1)^{R-r}\right]\; \; \;,
\label{eq:critic_cayley_app}
\end{equation}
suppressing the path that was followed to reach the node. From
Eq.(\ref{eq:common_frequency_cayley}) it is easily seen that when
$\Omega>0$ [$\Omega<0$], Eq.(\ref{eq:critic_cayley_app}) is always
negative [positive], monotone and increasing [decreasing] and
takes its minimum [maximum] value for $r=1$. Imposing the bound of
the sine function ( $\left|\sin{\left(\varphi\right)}\right|\leq
1$ ) to Eq.(\ref{eq:critic_cayley_app}), with $r=1$,  and
inserting Eq.(\ref{eq:common_frequency_cayley_app}) we obtain
\[
\left|\frac{\Delta\omega}{K}\right|_C = \frac{ 1}
{ z\sum_{q=0}^{R-2}\left(z-1\right)^q+\left(z-1\right)^{R-1} } + \left(1+\epsilon \right)\;\;\; .
\]
For $z=2$ again we obtain Eq.s(\ref{eq:critic_right}) and
(\ref{eq:critic_left}) with $s=N/2$. For $z>2$ we can again use
the truncated geometric series of Eq.(\ref{eq:truncated_series})
and obtain Eq.(\ref{eq:critic_cayley}).

\end{appendix}

\end{document}